\documentclass[useAMS,usenatbib]{mn2e}
\usepackage{graphicx}
\usepackage{amssymb,amsmath}
\usepackage[T1]{fontenc}
\usepackage{aecompl}
\title[Application of the Baade-Wesselink method to H254 in IC348]{Application of the Baade-Wesselink method to a pulsating 
cluster Herbig Ae star: H254 in IC348\thanks{Based on observations
  made with the Italian Telescopio Nazionale Galileo (TNG) operated on
  the island of La Palma by the Fundación Galileo Galilei of the INAF
  (Istituto Nazionale di Astrofisica) at the Spanish Observatorio del
  Roque de los Muchachos of the Instituto de Astrofisica de
  Canarias}}

\author[V. Ripepi et al.]{V. Ripepi$^{1}$\thanks{Please send offprint
    requests to {\it ripepi@oacn.inaf.it}},
R. Molinaro$^{1}$
M. Marconi$^{1}$,   
G. Catanzaro$^{2}$,
R. Claudi$^{3}$,
\and  
J. Daszy\'nska-Daszkiewicz$^{4}$, 
 F. Palla$^{5}$, 
S. Leccia$^{1}$, 
S. Bernabei$^{6}$
\\
$^{1}$INAF-Osservatorio Astronomico di Capodimonte, I-80131, Napoli, Italy\\
$^{2}$INAF-Osservatorio Astrofisico di Catania, I-95123, Catania, Italy\\
$^{3}$INAF-Osservatorio Astronomico di Padova, I-35122, Padova, Italy \\ 
$^{4}$Instytut Astronomiczny, Uniwersytet Wroc$\l$awski, ul. Kopernika
11, PL-51-622 Wroc$\l$aw, Poland \\
$^{5}$INAF-Osservatorio Astrofisico di Arcetri, I-50125, Firenze, Italy\\ 
$^{6}$INAF-Osservatorio Astronomico di Bologna, I-40127, Bologna, Italy\\ 
}

\begin{document}

\date{}

\pagerange{\pageref{firstpage}--\pageref{lastpage}} \pubyear{2002}

\maketitle

\label{firstpage}

\begin{abstract}

In this paper we present new photometric and radial velocity data
for the PMS $\delta$ Sct star H254, member of the young cluster 
IC 348. Photometric V,R$_C$,I$_C$ light curves were obtained at the
Loiano and Asiago  telescopes. The radial velocity data was acquired
by means of the SARG@TNG spectrograph.  
High-resolution spectroscopy allowed us to derive precise 
stellar parameters and the chemical composition of the star, obtaining: T$_{\rm
  eff}$\,=\,6750\,$\pm$\,150~K; $\log g$\,=\,4.1\,$\pm$\,0.4 dex; 
[Fe/H]=-0.07$\pm$0.12 dex.  Photometric and spectroscopic data were  
used to estimate the total absorption in the $V$ band
A$_{\rm V}$=2.06$\pm$0.05 mag, in agreement with previous
estimates. 
We adopted the technique of the difference
in phase and amplitude between different photometric bands and radial
velocities to verify that H254 is (definitely) pulsating in a radial mode. 
This occurrence allowed us to apply the CORS
realization of the Baade--Wesselink method to obtain 
a value for the linear radius of H254 equal to 3.3$\pm$0.7 $R_\odot$. 

This result was used in conjunction with photometry and effective
temperature to derive a  distance estimate of 273$\pm$23 pc for H254,
and, in turn for IC 348, the host cluster. 
This value is in agreement within the errors with the results derived 
from several past determinations and the evaluation obtained through
the Hipparcos parallaxes. 
Finally, we derived the luminosity of H254 and studied its position
in the Hertzsprung--Russell diagram. From this analysis it results
that this $\delta$--Scuti occupies a position close to the red edge of
the instability strip, pulsates in the fundamental mode, 
has a mass of about 2.2 $M_\odot$ and an age
of 5$\pm$1 Myr, older than previous estimates.
\end{abstract}

\begin{keywords}
stars: pre-main-sequence -- stars: variables: T Tauri, Herbig Ae/Be --
stars: variables: $\delta$ Scuti --
stars: fundamental parameters -- stars: abundances -- open clusters
and associations: individual: IC 348.
\end{keywords}

\section{Introduction}

Herbig Ae/Be stars \citep{herbig60} are intermediate-mass stars (1.5
M$_{\odot} <$ M $\lesssim$ 4-5 M$_{\odot}$) experiencing their 
pre-main-sequence (PMS)  phase. Observationally, these stars show: i) 
spectral type B, A or early F;
ii) Balmer emission lines in the stellar spectrum; iii) Infrared
radiation excess (in comparison with normal stars) due to
circumstellar dust. Additionally, often these objects show 
significant photometric and spectroscopic variability on very different
time-scales. Variable extinction due to circumstellar dust causes
variations on timescales of weeks, whereas clumps (protoplanets and
planetesimals) in the circumstellar disk or 
chromospheric activity are responsible for hours to days variability
\citep[see e.g.][]{Catala2003}.

It is now well established that intermediate-mass PMS stars during
contraction towards the main sequence cross the instability strip for   
more evolved stars. These young, pulsating
intermediate-mass stars are collectively called PMS $\delta$ Sct  and their
variability, similarly to the evolved $\delta$ Sct variables, is
characterized by short periods ($\sim$0.5 h$\div$5 h) and small amplitudes 
\citep[from less than a millimag to a few hundredths of magnitude, see, e.g.]
[and references therein]{kurtz95,ripepi03,ripepi06,zwintz08,ripepi11}.

The first theoretical investigation of the PMS $\delta$ Sct instability strip based
on nonlinear convective hydrodynamical models was carried out by
\citet{marconi98}. In this seminal paper, these authors calculated the
instability strip topology for the first three
radial modes and showed that the interior structure of
PMS stars crossing the instability strip is significantly different
from that of more evolved Main Sequence stars (with the same mass and
temperature), even though the envelope structures are similar. 

The subsequent theoretical works by several Authors \citep[see
e.g.][and references therein]{suran01,Marconi2004,ruoppo07,dicriscienzo2008,guenther2004} 
showed that, when a sufficient number of observed frequencies are
available for a single object (usually a mixture of radial and
non-radial modes), the asteroseismic interpretation of the data allows 
to estimate with good accuracy the  
position of these variables in the HR diagram and to get insights
into the evolutionary status of the studied objects.
From the observational point of view, multi-site campaigns
\citep[e.g.,][]{ripepi03,ripepi06,bernabei09} and space observations with MOST
and CoRoT \citep[e.g.,][for NGC2264]{zwintz09,zwintz11} were
progressively able to provide improved sets of frequencies to be 
interpreted at the light of asteroseismic theory. Eventually, the
comparison between theory and observations allows to 
obtain stringent constraints on the stellar parameters and the internal structure 
of these objects. 

On the other hand, the asteroseismic techniques are not suitable 
for monoperiodic pulsators, for which it is difficult to discern
whether the observed mode is radial or not, making it almost impossible to  
use stellar pulsation theory to constrain the stellar parameters of
the star.  
As demonstrated by various authors \citep[see e.g.][and references
therein]{jagoda03,jagoda05}, this problem can be faced by considering  
amplitude ratios and phase differences from multi-colour photometry
and comparing these with models. This procedure allows to estimate 
the value of the harmonic degree $\ell$. 
However, for monoperiodic pulsators, an additional possible test to
discriminate between radial and non-radial modes is potentially represented by the
so-called Baade-Wesselink test. As well known, the Baade-Wesselink
method combines photometric and radial velocity data along the
pulsation cycle producing as output an estimate of both the radius and the
distance of the investigated pulsating star \citep[see][for a
review]{gautschy87}. 
For radially oscillating stars, the line-of-sight motion
(responsible for the radial velocity curve) and the subtended area of
the star (responsible for the light curve) are in phase, so that the
method gives a reasonable estimate of the stellar radius.  This
occurrence does not hold in the case of non-radial pulsation.
In fact, for a $p$-mode with $l=2$, we would obtain a negative value
for the radius, while for $l=1$ the result would be an unrealistically
large radius \citep{Unno89}. 

In this context, among the already known PMS $\delta$ Scuti
variables, the IC 348 cluster member H 254 \citep{her98} is an ideal
candidate. Indeed, this pre-main sequence
star is fairly bright (V=10.6 mag) and it has already been found
by \citet[Paper I hereinafter]{rip02} to be a monoperiodic pulsator
with $f=7.406$ c/d (period$\approx$3.24 h) and a peak to peak $V$
amplitude of $\sim$0.02 mag. These results 
were confirmed by  \citet{kiz05} on the basis of completely
independent photometric observations. 

A comparison between observations and the radial analysis based on the 
models by \citet{marconi98} suggests that H254 could be a 2.6 $M_{\odot}$ or a 2.3 $M_{\odot}$
pulsating in the first overtone or in the fundamental mode
respectively. 
The application of the Baade-Wesselink method to this
pulsator would allow us to test this result.  Moreover, such an
analysis would
also provide an estimate of both the radius and the distance of
this star, and in turn, its position in the HR Diagram, as well as 
an independent estimate of poorly known distance to the parent 
cluster IC 348, the parent cluster, which is poorly known 
\citep[see][for a discussion on this topic]{her98}. 

At variance with the case of RR Lyrae and especially Classical Cepheids,
Baade-Wesselink application to $\delta$ Sct stars is extremely rare
in the literature, and concerned only with High Amplitude $\delta$ Sct
stars \citep[see e.g.][]{burky86}. This occurrence is probably due to the
limited use of $\delta$ Sct as distance indicators for extragalactic
objects, as well as to the availability of precise Hipparcos
parallaxes for several close objects. In addition, there is an
observational difficulty in 
obtaining high precision radial velocity measurements
with the short time exposures needed to avoid light curve smearing for
these fast pulsators. Hence in this paper we present the first 
attempt to apply the BW technique to a low
amplitude $\delta$ Sct star. 

The organization of the paper is as follows: in section 2 we present
the photometric and spectroscopic observations; in section 3 we
discuss the determination of the stellar parameters and the abundance
analysis; section 4 reports on the technique used to discriminate between 
radial and non-radial pulsation in H254; in section 5 we apply the
CORS method to derive the linear radius of our target; section 6
discusses the inferred distance and position in the HR
diagram; in section 7 we summarize the main achievements of this paper.

\begin{figure}
\centering
\includegraphics[width=9cm]{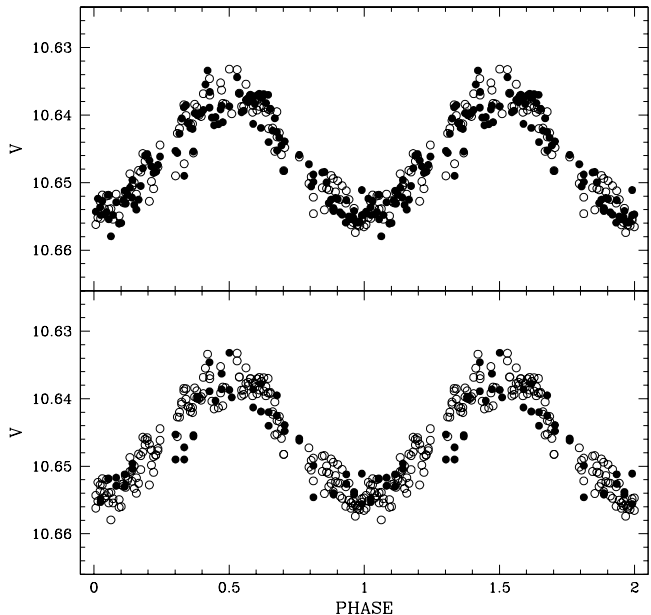}
\caption{Folded light curve in $V$ for H254. Top panel shows the light curves 
obtained using H83 (filled circles) and H89 (open circles) as
comparison stars, respectively. Bottom panel shows the light curved
obtained with the Asiago (filled circles) and Loiano (open circles)
telescopes, respectively.}  
\label{comparison}
\end{figure}

\section{Observations and data reduction}

The Baade-Wesselink method relies on both photometric and radial
velocity (RV) observations.
To minimize possible phase shifts between
light and  RV curves, that could represent a significant source of uncertainty in the application of the method, 
we aimed at obtaining photometric and spectroscopic observations as simultaneous as possible. 
This was achieved only partially, as shown in Tab.\ref{log} , where
the  log of the observations is shown. Indeed, due to the adverse weather conditions, we were able to gather useful 
photometric data during only two nights, one of which was luckily very close to the spectroscopic ones 
so that we are confident that our results are not
hampered by the above quoted possible phase shifts.

\begin{figure}
\centering
\includegraphics[width=9cm]{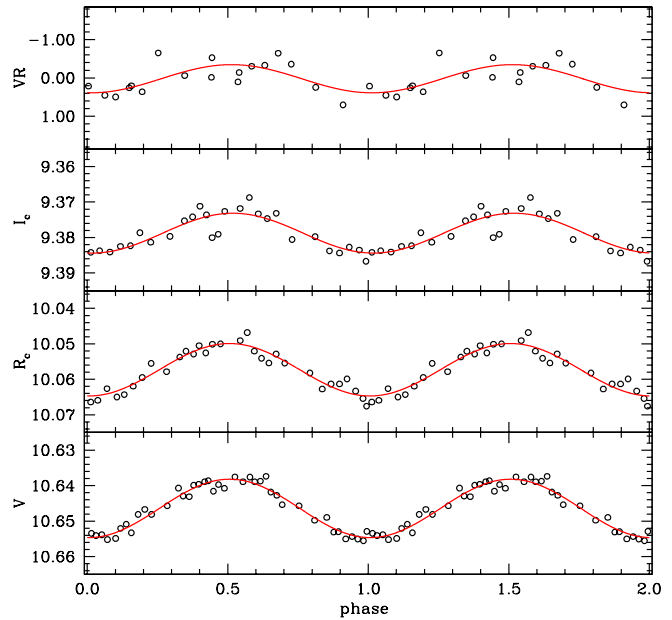}
\caption{Top panel: folded radial velocity curve of H254. Second to fourth
  panels show the smoothed folded light curves in $I_c$, $R$ and $V$,
  respectively. In all panels the solid line represent a sinusoidal
  least-squares fit to the data (see text)}  
\label{datiH254}
\end{figure}

\subsection{Photometry}

Photometric data were acquired in the Johnson-Cousins system $V,R_C$ and $I_C$ with the 
AFOSC\footnote{http://archive.oapd.inaf.it/asiago/2000/2300/2310.html} and 
BFOSC\footnote{http://www.bo.astro.it/loiano/observe.htm} instruments at the 1.8m 
Asiago and 1.54m Loiano telescopes, respectively.  The AFOSC@1.82m instrument was equipped with 
a TK1024AB 1024x1024 CCD, with a pixel size of 0.47$^{''}$ and a total field
of view of about 8.1$^{'}\times8.1^{'}$. The BFOSC@1.54m was
equipped with a EEV CCD 1300x1340 pixels of individual size
0.58$^{''}$, and a total field of view of 13$^{'}\times13^{'}$. 

The data were reduced following the usual procedures (de-biasing,
flat-fielding) and using standard IRAF\footnote{IRAF is distributed by the National Optical Astronomy Observatories,
which are operated by the Association of Universities for Research
in Astronomy, Inc., under cooperative agreement with the National
Science Foundation.} routines. 

To measure the light variations of H254, we adopted the differential
photometry technique. We know from Paper I that the stars H83 and H89
are isolated, constant, and bright enough to provide a very high S/N, and hence
are the ideal comparison stars. 

The aperture photometry was carried out using custom routines written in the 
MIDAS\footnote{http://www.eso.org/sci/software/esomidas/} environment
with apertures of about 17$^{''}$ for both data sets. 
To join the Asiago and Loiano observations, we have first calculated the differences in
average magnitude between the two set (of differential magnitudes) due to long term photometric
variation not caused by pulsation (see Paper I). The correction was of
the order of a few hundredths of magnitudes for all the filters. Due
to the largest data set, we retained the
Loiano photometry as reference and corrected the Asiago one. 

To secure the absolute photometric calibration we have first
calibrated the comparison stars H83 and H89, using the $VR_CI_C$ photometry 
by \citet{trullols1997} and then for each star and filter, we added
the magnitude differences with respect to H254. In
this way we obtained two distinct calibrated time series for each
band, as reported in Tab.~\ref{photometry},  which is published in its entirety in
the on-line version of the paper. 
The quality of the data at varying the comparison star or the telescope is
shown in Fig~\ref{comparison} in the case of the $V$ band, but similar
results are obtained for $R_C$ and $I_C$. Having verified that all
the data we gathered are compatible to each other, the next step was to
merge all the data sets (at varying the comparison star and the telescope) 
for each filter. We then smoothed the resulting light curves by
applying a box car-like smoothing algorithm, making it 
easier to perform a least-squares fit to the data. 
The smoothed light curves were then 
fitted with a Fourier series: 
\begin{equation}  
mag(\phi) = A_0+\sum_{i=1}^nA_i\sin(2\pi i\phi+\Phi_i).
\end{equation}
Typically only one harmonic (n=1 in the formula above) is sufficient
to describe the sinusoidal shape data of the target. The
goodness of the photometry is testified by the very low rms of the
residuals around 
the fits: $0.0015$ mag, $0.0018$ mag 
and $0.0025$ mag for $V,R_c$ and $I_c$ bands, respectively.

The result of all these steps is reported in the last three panels of
Fig.~\ref{datiH254}, where, as in Fig.\ref{comparison}, we have phased the photometry using
the period given in Paper I, and redetermined the epoch of the
minimum light (which was better defined with respect to the maximum) 
on the Loiano data set. 

\begin{table}
\caption{Log of the observations. Starting from the left, the
  different columns report: the observing mode; Heliocentric Julian Day (HJD) of start and end observations; the
  length of the time series; the exposure times (in $V,R,I_C$,
  respectively  for photometry). Note that HJD=HJD-2454000. 
} 
\label{log} 
\centering                
\begin{tabular}{cccccc}     
\hline
\noalign{\smallskip} 
Mode & Obs. & HJD-s & HJD-e& Length & Exptime \\
          &                     &           d   &        d      &       h       &          s \\ 
\noalign{\smallskip}
\hline
\noalign{\smallskip}  
Phot. & Loiano   &        415.308 & 415.437 & 3.1 & 12,9,8 \\  
Phot  & Asiago  &        434.395  & 434.547 & 3.6  &  10,5,6 \\
Spec. & TNG     &     431.695  & 431.759   & 1.5  & 1020 \\
Spec. & TNG     &     432.338 & 432.747 & 9.8  &  1020 \\
\noalign{\smallskip}
\hline
\noalign{\smallskip}
\end{tabular}
\end{table}

\begin{table}
\caption{$V,R_C,I_C$ photometry for H254. From left to right we 
report: HJD (in days); magnitude; filter; comparison star
used. Note that HJD=HJD-2400000. } 
\label{photometry} 
\begin{center}        
\begin{tabular}{cccc}     
\hline
\noalign{\smallskip} 
HJD    &  Mag & filter & Comparison \\
\noalign{\smallskip}
\hline
\noalign{\smallskip}   
 54434.39491  &  10.6400   &   V  &   H89  \\ 
 54434.40091  &  10.6389   &   V  &   H89  \\ 
 54434.40693  &  10.6386   &   V  &   H89  \\ 
 54434.41081  &  10.6387   &   V  &   H89  \\ 
 54434.42263  &  10.6413   &   V  &   H89  \\ 
 54434.42652  &  10.6419   &   V  &   H89  \\ 
 54434.43035  &  10.6440   &   V  &   H89  \\ 
 54434.43444  &  10.6425   &   V  &   H89  \\ 
 54434.43831  &  10.6439   &   V  &   H89  \\ 
\noalign{\smallskip}
\hline
\noalign{\smallskip}
\end{tabular}
\end{center}
Table~\ref{photometry} is published in its entirety only in the
electronic edition of the journal. 
A portion is shown here for guidance regarding its form and content.
\end{table}

\begin{table}
 \caption{Observation log for spectroscopic data. 
The B star (HD5394) as well as the target template (the spectrum without the iodine cell)
  were observed during each night to determine the best instrumental profile modelling.} 
\label{TabSpectroscopy} 
\centering                
\begin{tabular}{ccccc}     
\hline                  
\hline            
Star  & Nr. spectra & Iodine cell & T$_{exp}$[s]&S/N \\
\hline       
HD5394  &2  & Yes & 180 & $>$ 200 \\
HD5394  &4  & No  & 60 & $>$ 200 \\  
HD5394  &4  & No  & 30 & $>$ 200 \\
H254    &6  & No  & 1800& 40 -- 45 \\
H254    &22 & Yes & 1020& 19--34 \\
\hline      
\end{tabular}
\end{table}



\subsection[]{Spectroscopic observations and radial 
velocity determination}

The spectroscopic observations were carried out with the SARG
instrument, which is a high-resolution (from $R=29000$ to $R=164000$) cross
dispersed echelle spectrograph covering a spectral range from
$\lambda$ = 370 nm to 1000 nm \citep[see][for details]{gratton2001}
mounted at Telescopio Nazionale Galileo
(TNG, La Palma, Canarie, Spain)\footnote{Note that SARG was dismissed 
  at TNG and replaced
  with the HARPS-N spectrograph.} .  

Using the yellow grism (spectral range $462-792$ nm) it is possible to insert in the light path a $I_2$-cell
and have, in this way, a superimposed (and stable) wavelength
reference on the spectrum of the star useful for accurate stellar
Doppler shifts measurements.  The majority of the spectra in our
data sample were acquired in this configuration with a resolution of
$R=164000$ during two observing nights (see table \ref{log}). A few
spectra of the target were acquired without the $I_2$-cell for
calibration purposes. The
reduction of the spectra (bias subtraction and flat fielding
correction) was performed using the common IRAF package facilities.
To unveil the radial velocity information
hidden in the star spectra, we need to reconstruct the spectra of the star together with 
the superimposed $I_2$ spectrum, using the measured instrumental
profile, a very high resolution $I_2$ spectrum and a high resolution 
and high signal to noise spectrum of the star.
To model the instrumental profile, it is necessary to obtain a
spectrum for a fast rotating B star acquired with and without the
iodine cell. A detailed description of the spectra available is shown 
in Tab.~\ref{TabSpectroscopy}.  

Besides, using the high resolution $I_2$ spectra we obtain the
instrumental profile deconvolving it by the B star spectrum. 
After that the observed spectrum of the star is compared with
a modeled spectrum of the star using all obtained elements. 

The reconstruction of the spectra and the comparison with the observed one
are performed using the AUSTRAL code by \citet{endl2000}. 
This code  takes in account the
instrumental profile changes among the spectrum subdividing it in
80-120 pixel chunks. For each of these
chunk a radial velocity is measured. Usually the spectrum
is subdivided in 400-600 chunks to which correspond a radial
velocity measurement. The final radial velocity measure
is obtained by the mean and standard deviation of all the
results of the chunks. The results of the measurement are
reported in Fig.~\ref{datiH254} and Tab.~\ref{rv}.

\begin{table}
\caption{Radial velocities for H254 measured with SARG.  
Note that HJD=HJD-2400000. } 
\label{rv} 
\centering                
\begin{tabular}{ccc}     
\hline
\noalign{\smallskip}  
HJD    &  RV & $\sigma_{\rm RV}$ \\
     d   &   Km/s      &    Km/s   \\ 
\noalign{\smallskip}
\hline
\noalign{\smallskip}    
54431.69459  &      0.246  &      0.235       \\
 54431.70765  &      0.705  &      0.327      \\
 54431.72033  &      0.214  &      0.194      \\
 54431.73337  &      0.495  &      0.322      \\
 54431.74605  &      0.361  &      0.495      \\
 54432.40098  &      0.452  &      0.257      \\
 54432.41366  &      0.205  &      0.300      \\
 54432.42647  &     -0.652  &      0.490      \\
 54432.43916  &     -0.061  &      0.442      \\
 54432.45197  &     -0.015  &      0.324      \\
 54432.46467  &      0.103  &      0.412      \\
 54432.47749  &     -0.330  &      0.354      \\
 54432.49018  &     -0.359  &      0.395      \\
 54432.54718  &      0.258  &      0.340      \\
 54432.60579  &     -0.305  &      0.378      \\
 54432.61846  &     -0.645  &      0.347      \\
 54432.72134  &     -0.530  &      0.454      \\
 54432.73436  &     -0.142  &      0.313      \\ 
\noalign{\smallskip}
\hline
\noalign{\smallskip}
\end{tabular}
\end{table}

\section{Spectral type and atmospheric parameters}

The same spectroscopic data used to build the reference spectrum for radial velocity 
determination can be used to estimate the 
intrinsic stellar parameters for the target star.

\subsection{Determination of effective temperature}
\label{teff}
Any attempt devoted to a detailed characterization of the chemical abundance
pattern in stellar atmospheres is strictly linked with the accuracy of
effective temperature and surface gravity determination.

In this study, we derived the effective temperature by using the ionization
equilibrium criterion. In practice, we adopted as T$_{\rm eff}$ the value
that gives the same iron abundance as computed from a sample spectral lines
both neutral and in first ionization stage.

First of all, we selected from the TNG spectrum a sample of iron lines
with the principal requirement that they do not show any sensitivity to 
the gravity (this condition has been checked using spectral
synthesis) i.e. synthetic lines computed for a wide range of gravities
did not show any appreciable variations.
We found six lines belonging to Fe{\sc i} and only two to 
Fe{\sc ii}. The list of these lines with their principal atomic parameters 
is reported in Tab.~\ref{fe_lines}. 

For each line, the equivalent width and the central wavelength have been measured with 
a Gaussian fit using standard IRAF routines. As the main source of errors in
the equivalent width measurements is the uncertain position of the continuum,
we applied the formula given in \citet{leo95}, which takes
into account the width of the line and the rms of the continuum. According
to the S/N of our spectrum and to the rotational velocity of our target 
($v \sin i$\,=\,85\,$\pm$\,2~km~s$^{-1}$, see next section),
we found that the error on the equivalent widths is $\approx$\,20 m{\AA}.

Then, once fixed the observed equivalent widths, we computed for each spectral line
the theoretical curves in the log Fe/N$_{\rm tot}$~-~T$_{\rm eff}$ plane.
Since our target is reported in the literature as a F0 star (SIMBAD database), 
we explored the range in temperature between 6500~K and 7000~K. Calculations 
have been performed in two separate steps:

\begin{itemize}

\item use of ATLAS9 \citep{kur93} to compute the LTE 
      atmospheric models 

\item use of WIDTH9 \citep{kur81} to derive abundances 
      from single lines.

\end{itemize}

\begin{figure}
\centering
\includegraphics[width=9cm]{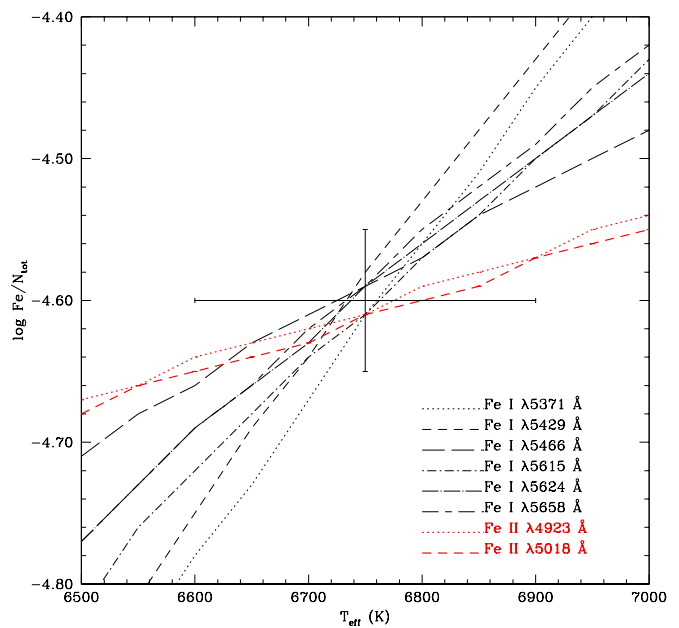}
\caption{Behaviour of iron abundances as a function of effective temperature. The black 
curves refer to neutral iron, while the red ones to single
ionized.}
\label{ab_teff}
\end{figure}

The locus in common with all the curves showed in Fig.~\ref{ab_teff} 
represents the effective temperature and iron abundance of H254. To give a realistic
estimation of the errors on these values, we repeated the same calculations as before, but
varying the EW by a quantity equal to its experimental error. In practice the value of 
T$_{\rm eff}$ and $\log N_{\rm Fe}/N_{\rm Tot}$ derived for EW\,$\pm$\,$\delta$\,EW gave
us the extension of the error bars. Finally, we adopted T$_{\rm eff}$\,=\,6750\,$\pm$\,150~K
and $\log N_{\rm Fe}/N_{\rm Tot}$\,=\,$-$4.60\,$\pm$\,0.05.

\begin{table}
\caption{List of iron lines used for temperature determinations, $\log gf$ and
relative reference are also reported.} 
\label{fe_lines} 
\centering                
\begin{tabular}{ccc}     
\hline                  
\hline            
Sp. line & $\log gf$ & Reference \\
\hline       
Fe{\sc i} $\lambda$5371.489 & $-$1.644 & \citet{nist08} \\
Fe{\sc i} $\lambda$5429.696 & $-$1.879 & \citet{nist08} \\
Fe{\sc i} $\lambda$5466.396 & $-$0.630 & \citet{castelli04} \\
Fe{\sc i} $\lambda$5615.644 & $-$0.615 & \citet{castelli04} \\
Fe{\sc i} $\lambda$5624.542 & $-$0.900 & \citet{castelli04} \\
Fe{\sc i} $\lambda$5658.816 & $-$0.920 & \citet{castelli04} \\
Fe{\sc ii} $\lambda$4923.927 & $-$1.320 & \citet{nist08} \\
Fe{\sc ii} $\lambda$5018.440 & $-$1.210 & \citet{nist08} \\
\hline      
\end{tabular}
\end{table}

\subsection{Determination of surface gravity}
\label{grav}
For early F-type stars, one of the method commonly used in literature for
the determination of the luminosity class of stars is to estimate the
strength of the ratios between the spectral blend at 
$\lambda \lambda$ 4172-4179 {\AA}, mostly Fe{\sc ii} and Ti{\sc ii} lines,
and the one at $\lambda$4271 {\AA}, mostly composed of Fe{\sc i} lines.
\citep{gray89}.

Since the TNG spectra do not cover the spectral region of interest for 
this purpose, we used the Loiano telescope spectra. We computed the 
$\lambda \lambda$ 4172-4179/4271 ratio which resulted 1.5\,$\pm$\,0.2,
roughly corresponding to a luminosity class of V \citep{gray89}.

To go deeply in detail and to derive the surface gravity of our target,
we computed the theoretical behaviour of that ratio as a function of $\log g$.
After having fixed the T$_{\rm eff}$ to the value found in Sect.~\ref{teff},
we computed ATLAS9 atmospheric models with gravities spanning the range
between 3.5 and 4.5 dex. By using this curve, we converted our measured
ratio in a measurement of gravity, obtaining: $\log g$\,=\,4.1\,$\pm$\,0.4.


Finally, considering the results of the Sects.~\ref{teff}-\ref{grav},
we conclude that the spectral type of our target is F3~V.

\subsection{Abundance analysis}

To derive chemical abundances, we undertook a synthetic modeling of the
observed spectrum. The atmospheric parameters inferred in the previous
sections have been adopted to compute an ATLAS9 LTE model
with solar ODF. This model has been applied to SYNTHE code \citep{kur81}
to compute the synthetic spectrum. 

\begin{figure}
\centering
\includegraphics[width=9cm]{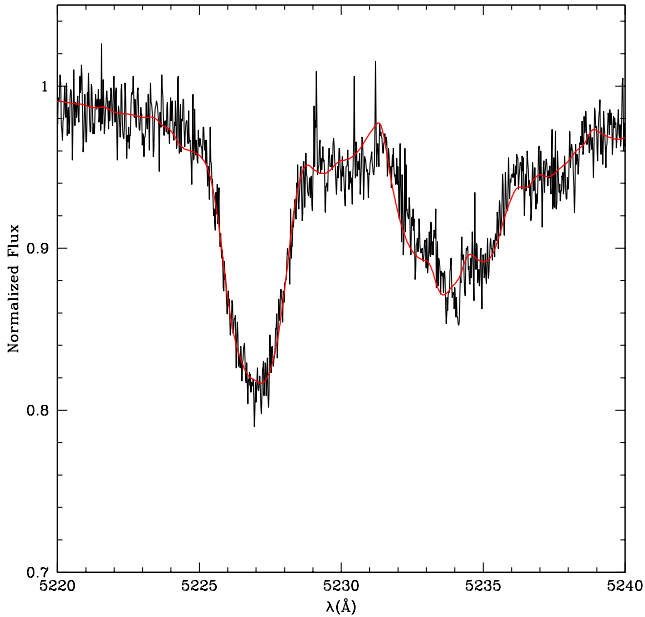}
\caption{Portion of spectra modeled with $v \sin i$\,=\,85~km~s$^{-1}$.
The spectral lines are: Fe{\sc i}~$\lambda$5227 {\AA} and 
Fe{\sc ii}~$\lambda$5233 {\AA}.}
\label{fit}
\end{figure}

The rotational velocity of H254 has been derived by matching metal lines with 
synthetic profiles. The best fit occurred for 85\,$\pm$\,2~km~s$^{-1}$, where
the error has been estimated as the variation in the velocity which increases the 
$\chi^2$ of a unit. An example of the goodness of our fit is showed in Fig.~\ref{fit}.

Practically, we divided all the spectral range covered by our data in a number 
of sub-intervals $\approx$~100~{\AA} wide. For each interval we derived the 
abundances by a $\chi^2$ minimization of the 
difference between the observed and synthetic spectrum.
Line lists and atomic parameters used in our modeling are from 
\citet{kur95} and the subsequent update by \citet{castelli04}. The iron abundance
found in Sect.~\ref{teff} has been used as guess input to speed up the calculations.
The adopted iron abundance with its error is the one reported in Tab.~\ref{ab}.

\begin{figure}
\centering
\includegraphics[width=9cm]{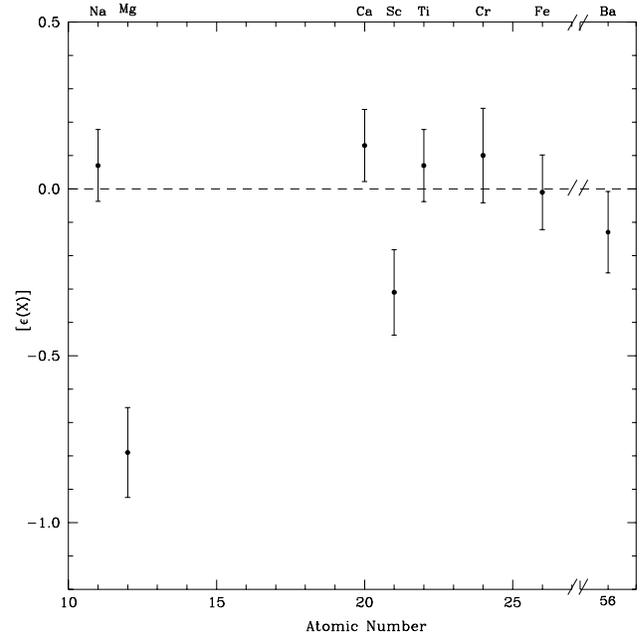}
\caption{Chemical pattern derived for H254, the horizontal dashed line 
represents the solar abundances.}
\label{pattern}
\end{figure}

In Table~\ref{ab} and Fig.~\ref{pattern} we report the abundances derived in our analysis
expressed in the usual logarithmic form relative to the total number of 
atoms $N_{\rm Tot}$. To easily compare the chemical pattern of H254 with the
Sun, we reported in the last column the differences with the solar values
as taken from \citet{asplund05}. Errors reported in Table~\ref{ab} for a 
given element are the standard deviation on the average computed among the 
various abundances determined in each sub-interval. When a given element 
appears in one or two sub-intervals only, the error on its abundance evaluated 
varying temperature and gravity in the ranges 
$[ T_{\rm eff}\,\pm\,\delta T_{\rm eff}]$ and 
$[\log g\,\pm\,\delta \log g]$ is typically 0.10 dex.

Inspection of  Fig.~\ref{pattern} suggests an almost standard 
atmosphere, with the exception of a moderate under abundance ($\approx$~0.8~dex)
of magnesium and a slight under abundance ($\approx$~0.3~dex) of scandium.

\begin{table}
\caption{Chemical abundances derived for the atmosphere of H254. For each
element we report its abundance in the form $\log N_{\rm el}/N_{\rm Tot}$,
and the difference with respect the solar values \citep{asplund05}.} 
\label{ab} 
\centering                
\begin{tabular}{crr}     
\hline                  
\hline            
Elem. & $\log N_{\rm el}/N_{\rm Tot}$ & [$\epsilon (x)$]~~~~ \\
\hline       
Na & $-$5.80 $\pm$ 0.10 &    0.07 $\pm$ 0.10 \\
Mg & $-$5.30 $\pm$ 0.10 & $-$0.79 $\pm$ 0.13 \\
Ca & $-$5.60 $\pm$ 0.10 &    0.13 $\pm$ 0.11 \\
Sc & $-$9.30 $\pm$ 0.10 & $-$0.31 $\pm$ 0.13 \\
Ti & $-$7.07 $\pm$ 0.09 &    0.07 $\pm$ 0.11 \\
Cr & $-$6.30 $\pm$ 0.10 &    0.10 $\pm$ 0.14 \\
Fe & $-$4.60 $\pm$ 0.10 & $-$0.07 $\pm$ 0.11 \\
Ba & $-$10.00 $\pm$ 0.10 & $-$0.13 $\pm$ 0.12 \\
\hline      
\end{tabular}
\end{table}

\subsection{Reddening determination}\label{sec-redd}

For the following analysis it is important to evaluate 
 the interstellar absorption in the direction of H254. 
To this aim, we firstly calculated the mean magnitudes in $V,R_c$ and $I_c$
bands on the basis of the fitting curves illustrated in Fig.~\ref{datiH254}. 
Then we calculated the colours:  $(V-R_c)$=0.589 mag and
$(V-I_c)$=1.268 mag. Taking  advantage of the $T_{\rm eff}$ value 
estimated from spectroscopy, we can now compare  the 
observed $(V-R_c)$ and $(V-I_c)$ values with those tabulated by 
\citet[][their Tab.A5]{ken95} for the estimated $T_{\rm eff}$ of
H254.  
As a result we find that the tabulated colours are: 
$(V-R_c)$=0.240 mag and $(V-I_c)$=0.480 mag. Hence, the derived values of
$E(V-R_c$) and $E(V-I_c$) are, respectively, 0.349 mag and 0.788
mag. 

The uncertainty  on these reddening values has been estimated by
analyzing how the theoretical colors change according to the error on
the spectroscopic temperature. The resulting uncertainties are $\delta
E(V-R_c)=0.015$ mag and $\delta E(V-I_c)=0.025$ mag. 

From the derived reddening values, we calculated the absorption A$_V$
as the weighted mean of the two values A$_V$=5.956$E(V-R_c$) and
A$_V$=2.612$E(V-I_c$), where the numerical coefficients are obtained
from \citet{car89}  assuming R$_V$=3.1. The value we use in this work
is A$_V$=2.06$\pm$0.05 mag.

\section{Mode identification by means of phase/amplitude difference
analysis}
\label{modePulsation}

\begin{figure*}
\centering
\includegraphics[width=17cm]{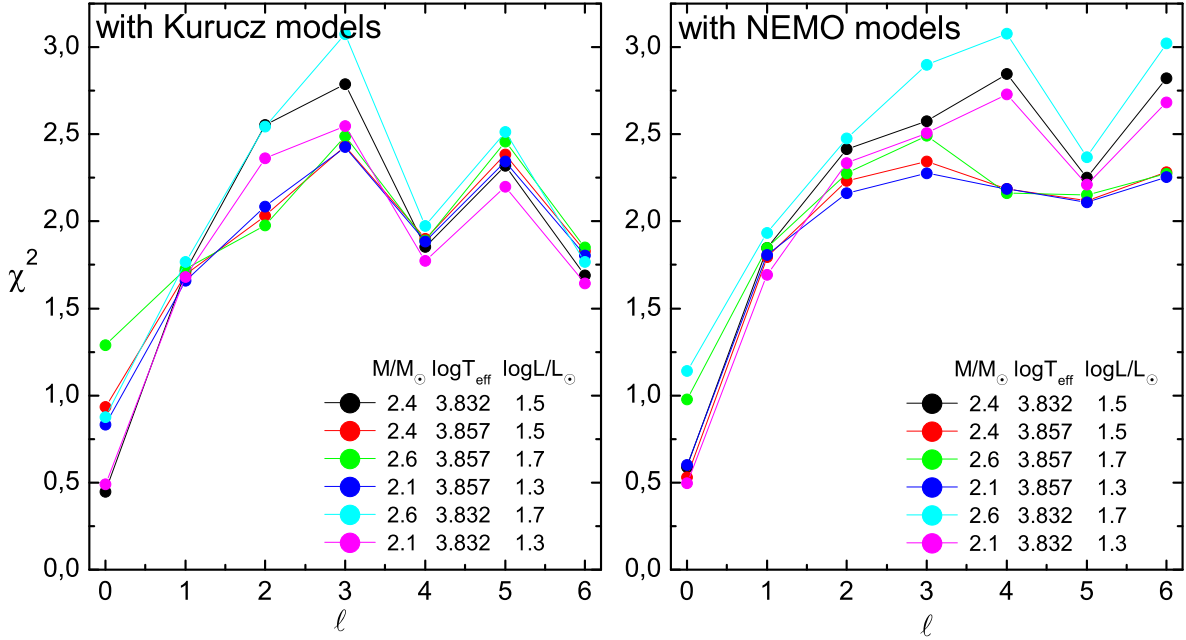}
\caption{The discriminant, $\chi^2$ as a function of the mode degree, $\ell$, for
pulsation frequency $\nu$ = 7.406 c/d of H254 as obtained from the fit
of the photometric amplitudes and phases. The five lines correspond to models
consistent with the observations. Two stellar model atmospheres were considered:
the Kurucz models (left panel) and The NEMO models (right panel). }  
\label{Jagoda}
\end{figure*}

Identification of the mode degree, $\ell$, for the detected pulsation frequency $\nu=7.406$ c/d of H254
has been done using the method of \citet{jagoda03,jagoda05}.
In this method the effective temperature perturbation, measured by the complex parameter $f$,
is determined from observation instead of relying on the linear non adiabatic computations
of stellar pulsations. The advantage is that we can avoid in that way uncertainties resulting
from theoretical modelling, e.g., an impact of subphotospheric convection on pulsation.
To apply the quoted method, the $VR_CI_C$ photometric data presented in
Sect. 2 are not well suited  because of the poor time
sampling. Therefore, we adopted the $uvby$ Str\"omgren photometry 
published in Paper I\footnote{We actually published only
  $b,y$ photometry in Paper I. Here we use also the data in the $u,v$
  bands. The $uvby$ light curves are available upon request.}.
In Fig.~\ref{Jagoda} we show the discriminant $\chi^2$ as a function
of $\ell$, as determined from the fitting of the theoretical amplitudes and phases to the observational values
in all $uvby$ passbands, simultaneously.
To illustrate how robust is the method to uncertainties
in stellar parameters (e.g., effective temperature, luminosity and mass) we considered five models
located in the error box of H254.

Moreover, two models of stellar atmospheres were adopted. In the left
panel of Fig.~\ref{Jagoda} we show results obtained with
the Kurucz models \citep{Kurucz04}, whereas the right panel of Fig.~\ref{Jagoda}
reports the same analysis based on the NEMO models \citep{Nendwich04}.
As we can see, in both cases identification of $\ell$ is clear: the pulsational frequency of H254
is associated with the radial mode.

\section{Baade--Wesselink application to H254}

In section~\ref{modePulsation} we have show that H254 pulsate in a 
radial mode. In this section we aim at applying to our target a
particular realization of the Baade--Wesselink method called CORS \citep{cac81} to 
derive the linear radius, $R$, of this low amplitude $\delta$--Scuti 
star. 
To date, the CORS method has been applied only to Cepheid stars
\citep[see][and references therein]{mol12},
characterized by light curves with large amplitude (typically $\sim$ 1 mag in the
visible) with respect to H254 ($\sim$0.02 mag). However, considering
that: i) the physics governing the pulsation in H254 is the same as in
Cepheids and ii) the light curve of H254 seems to be absolutely
smooth and not showing any kind of shock (as it happens e.g. for RR
Lyrae stars), hence we judge that we can safely apply the CORS method
to our target object.  This is done in the following section.

\subsection{The CORS equation}

The basic equation of the CORS method can be easily derived
starting from the surface brightness: 
\begin{equation}
S_V=m_V+5\log\theta
\label{eq-sv}
\end{equation}
where $\theta$ is the angular diameter (in mas) of the star and $m_V$
is the apparent magnitude in the V band. If we differentiate
eq.(\ref{eq-sv}) with respect to the pulsational phase ($\phi$),
multiply the result by the generic color index ($C_{ij}$) and
integrate along the pulsational cycle, we obtain:
\begin{equation}
q\int_{0}^{1}\!\ln
\Big\{R_0-pP\int_{\phi_0}^{\phi}\!v(\phi')d\phi'\Big\}C'_{ij}d\phi-B+\Delta
B=0
\label{eq-cors}
\end{equation}
where $q=\frac{5}{\ln 10}$, $P$ is the period, $v$ is the radial velocity 
and $p$ is the projection factor, which correlates radial and
pulsational velocities according to $R'(\phi)=-p\cdot P \cdot
v(\phi)$. The last two terms, $B$ and $\Delta B$, are given by: 
\begin{eqnarray}
B=\int_{0}^{1}C_{ij}(\phi)m'_V(\phi)d\phi \hspace{10pt} \\
\Delta B=\int_{0}^{1}C_{ij}(\phi)S'_V(\phi)d\phi \; . 
\label{eq-db}
\end{eqnarray}
and are connected to the area of the loop described during pulsational
cycle of the star, respectively, in the $V$--$C_{ij}$ plane and
$S_V$--$C_{ij}$ plane, typically with $B$ larger than $\Delta B$.

Equation (\ref{eq-cors}) is an implicit equation in the unknown radius
$R_0$ at an arbitrary phase $\phi_0$. The radius at any phase $\phi$
can be obtained by integrating the radial velocity curve. The main 
problem in the solution of eq.(\ref{eq-cors}) is the estimation of the
term $\Delta B$ which contains the unknown surface brightness. In the
case of Cepheids the term $\Delta B$, typically, assumes small values
with respect to the $B$ term and, consequently, in the original works it was
neglected \citep{cac81, onn85}. However recently \citet{mol11} found
that the Cepheid radii estimated by including the $\Delta B$ term are
more accurate than those based on the original version of the CORS
method. 

\citet{rip97} applied the CORS method to Cepheids and calculated the
$\Delta B$ term by calibrating the surface brightness expressed as
function of the $(V-R_C$) color index through the relation from
\citet{Gieren1989}. In more recent works by  \citet{Ripepi00,ruo04}  
and \citet{mol11, mol12}, the term $\Delta B$ has been calculated for a sample
of Cepheids by expressing it as a function of two colors using grids
of models. In the present work we calibrated the surface brightness 
using the relations given by \citet{ker08} and taking into account the
definition in eq.(\ref{eq-sv}). In particular, we defined the mean
surface brightness using the relations in $(V-R_c)$ and $(V-I_c)$ color indices:  
\begin{equation}
\begin{array}{l}
S_V=\frac{5}{2}(0.5100+1.2159(V-R_c)-0.0736(V-R_c)^2 + \\
 +0.4992+0.6895(V-I_c)-0.0657(V-I_c)^2)
\label{eq-our-sv} 
\end{array}
\end{equation}
while the color $C_{ij}$ in the eq.(\ref{eq-cors}) is chosen to be
$(V-I_c)$. We note here that the solution of the CORS method is not
dependent from the reddening because the first term of the
eq.(\ref{eq-cors}) contains the derivative of the color $C_{ij}$ and
the two terms $B$ and $\Delta B$ are connected with the area of the
loops described by the pulsating star. 

\subsection{The linear radius and the angular diameter}
To solve the CORS equation we used the fitted light curves and
interpolated the radial velocity data using the same procedure of 
photometry (see Fig.~\ref{datiH254}). The rms of residuals of the data
around the fitted radial velocity curve is 0.3 km/s. 
According to our result any phase shift between the
minimum of radial velocity and the maximum light is lower than 0.1
times the period,
in agreement with other authors \citep[see e.g.][]{bre76, nos92}. 

The mean radius obtained from the CORS method is equal to
$R=3.3 R_\odot$. To estimate the radius uncertainty we  performed
Monte Carlo simulations following the same procedure by
\citet{cac81}. These consist in varying all data points of light and
radial velocity curves using random shifts extracted from a Gaussian
distribution. The rms of the distribution in the case of photometry is
equal to the error on the data points (typically 0.001 mag), while for
the radial velocity it is equal to 10\% of the amplitude of the radial
velocity itself.  We have performed 5000 simulations and for each
simulated photometry and radial velocity curves the CORS equation has
been solved. Finally, the uncertainty has been obtained from the rms
of the resulting simulated radius distribution (clipped at 2-$\sigma$
to exclude possible outliers) and is equal to 0.7$R_\odot$. 
 
The angular size in mas, $\theta$, of H254 was obtained from the
relations provided by \citet{ker08} using $(V-R_c)$ and $(V-I_c)$ colors:
\begin{equation}
\log \theta =0.5100+1.2159(V-R_c)-0.0736(V-R_c)^2-0.2V
\end{equation}
\begin{equation}
\log \theta =0.4992+0.6895(V-I_c)-0.0657(V-I_c)^2-0.2V
\end{equation} 
with uncertainty of 4.5\% and 5.6\%, respectively. The derived mean angular
diameter is equal to $\theta_1=0.120\pm 0.005$ mas, for the color $(V-R_c)$, and
$\theta_2=0.088\pm 0.005$ mas for the color $(V-I_c)$. Hereafter we will
use the mean value $0.5(\theta_1+\theta_2)= 0.104$ mas
and, because of the evident inconsistency of the two previous measures,
we have conservatively estimated the uncertainty as their half
differences $0.5\left |\theta_1-\theta_2\right |=0.016$ mas. To
investigate the dependence of the angular diameter from the reddening,
we varied it within the uncertainty reported above and recalculated
the angular diameter. The resulting variations ($\sim 0.001$ mas) are
included into the error on the mean angular diameter.

\section{Results}

In this section we derive the distance of H254 and use it to estimate its
luminosity. Then, using the spectroscopic effective temperature, we
place H254 in the Hertzsprung--Russell diagram to estimate its mass
and age using theoretical evolutionary tracks and isochrones from
\citet{tog11}. 

\subsection{The distance of H254}

The distance of H254 can be obtained by using the simple equation
$d(pc)=\frac{9.304R(R_\odot)}{\theta(mas)}$, where R is the linear
radius derived from the Baade--Wesselink method, $\theta$ is the
angular diameter, obtained by using the relations by \citet{ker08},
and the constant factor takes into account the units of measure. 
The distance derived in this way is equal to 295$\pm$77 pc.

Assuming the validity of the Stefan--Boltzmann law $L=4\pi R^2\sigma
T_e^2$, where $\sigma$ is the Stefan--Boltzmann constant and $T_e$ is
the effective temperature of the star, the distance can be also
calculated from the following equation:
\begin{equation}  
d=10^{0.2(V-A_V+BC_V-M_\odot^{bol}+10\log\frac{T}{T_\odot}+5\log\frac{R}{R_\odot}+5)}
\end{equation}
where $BC_V$ is the bolometric correction from \citet{ken95},
$M_\odot^{bol}=4.64$ mag is the absolute bolometric magnitude of the Sun
 \citep{sch82}, $T_\odot=5777$ K is the Sun effective temperature. The
previous equation gives a distance value equal to 262$\pm$54 pc.
The uncertainty on the previous distance value was obtained from the
usual rules of propagation of errors.  The value obtained by the
solution of the CORS equation is in good agreement with that
obtained from the Stefan--Boltzmann law.  

A weighted average of the two distance estimates reported above gives 
273$\pm$23 pc, which represents our best estimate for the distance of
the H254.


We tested the consistency of this result with the distance
obtained by using the Period--Luminosity relation for
$\delta$--Scuti stars. Assuming that H254 pulsate in the fundamental
mode (this assumption is justified in the next section),  using the
PL relation recently derived  by \citet{mcn11}:

$$M_V=(-2.90\pm 0.05)\log P -(0.19\pm 0.15)[Fe/H]-(1.27\pm 0.05)$$ 

\noindent and adopting our metallicity estimate for H254, we obtain an absolute magnitude
$M_V=1.26\pm 0.07$ mag. By combining this value with the apparent visual
magnitude, $m_V=10.646$ mag, obtained from the fitting procedure of
the data and the extinction $A_V=2.06$ mag, derived in
Sect.~\ref{sec-redd},  we obtain a distance equal to 292$\pm$15 pc, in
very good agreement within the errors with the value from the BW
analysis.  


In the literature there are other estimates of the distance of H254
and/or the cluster IC 348 harbouring it. \citet{her98} reviewed all the
results till 1998 ranging from low values such as 240$^{128}_{-84}$ pc
\citep{trullols1997} and 260$\pm$16 pc \citep{cer93} to larger values,
namely 316$\pm$22 pc \citep{strom74}. In the end \citet{her98} 
assumed a distance of 316 pc (no error was given). \\
Subsequent investigations made use of the Hipparcos parallaxes.  
On the basis of 9 bright members of the cluster, \citet{sch99} derived
a distance of 260$\pm$25 pc. Contemporaneously, \citet{dezeeuw99} 
estimated the distance of the Per OB2 association, in which IC 348 is 
supposed to be embedded, and found a value of 318$\pm$27 pc. 
Assuming that there were no mistakes due to the inclusion of non-members
to the cluster and/or to the OB association, this discrepancy could mean that the Per
OB2 association and IC 348 are not at the same distance. To check this 
result, we recalculated the distance of IC 348 using the membership
criterion by \citet{sch99}, but adopting the revised Hipparcos parallaxes by 
\citet{vanleeuwen07}. The stars selected in this way are listed in Tab.~\ref{hipp}
together with the relevant information coming from the satellite. The
resulting distance is 227$\pm$25 pc, a result which tends to support a
lower value for the distance of IC348. This is confirmed even if we
include in the calculation only the
four stars closer than 30 arcmin to the center of the cluster, obtaining  
251$\pm$50 pc.   

Concluding, our estimate of 273$\pm$23 pc is in agreement within the
errors with the results derived from the  Period--Luminosity relation for
$\delta$--Scuti stars by \citet{mcn11} and those coming from the
Hipparcos parallaxes by \citet{sch99} as well as with the recalculation based
on the revised Hipparcos parallaxes by \citet{vanleeuwen07}. We are
only marginally in agreement with the larger value of the distance,
i.e. 310-320 pc found by \citet{her98} and \citet{dezeeuw99}.

\begin{table*}
\caption{Data used to recalculate the Hipparcos distance to IC
  348. From left to right the different columns show the
  identification of the star, the distance from the center of IC348,
  the RA and DEC, the new parallaxes by \citet{vanleeuwen07}, the
  proper motions in RA and DEC.} 
\label{hipp} 
\begin{tabular}{ccccccc}     
\hline
\noalign{\smallskip} 
name  & d               & RA  &   DEC       & $\pi$      & $\mu$$_{\rm RA}$ & $\mu$$_{\rm DEC}$ \\
          &  arcmin      & J2000  & J2000  &  mas   & mas   & mas   \\
\noalign{\smallskip}
\hline
\noalign{\smallskip}  
HIP 17465 & 0.05   &  03 44 34.187  &  +32 09 46.14   &    6.58$\pm$4.09    &    3.49$\pm$10.67   &    -4.86$\pm$14.96     \\ 
HIP 17448 & 8.13   &  03 44 19.132  &  +32 17 17.69   &    2.91$\pm$0.73    &    8.18$\pm$ 0.70   &   -10.43$\pm$0.69     \\ 
HIP 17561 & 21.57  &  03 45 39.164  &  +32 26 24.17   &    6.04$\pm$1.59    &   10.20$\pm$1.78   &    -4.87$\pm$1.64     \\ 
HIP 17631 & 27.89  &  03 46 40.871  &  +32 17 24.68   &    4.99$\pm$0.62    &    8.53$\pm$0.61   &    -8.29$\pm$0.57     \\ 
HIP 17845 & 58.12  &  03 49 07.301  &  +32 15 51.37   &    5.29$\pm$0.77   &    10.23$\pm$0.81  &     -8.82$\pm$0.74    \\ 
HIP 17113 & 60.67  &  03 39 55.685  &  +31 55 33.20   &    5.14$\pm$1.06   &     4.11$\pm$1.14  &     -4.60$\pm$0.92    \\ 
HIP 17052 & 69.80  &  03 39 22.934  &  +32 33 24.39   &    4.86$\pm$1.22   &     6.52$\pm$1.48  &    -16.53$\pm$1.15    \\ 
HIP 17966 & 83.50  &  03 50 27.513  &  +31 33 13.00   &    4.57$\pm$0.91   &     8.22$\pm$1.06  &     -6.72$\pm$0.90    \\ 
\noalign{\smallskip}
\hline
\noalign{\smallskip}
\end{tabular}
\end{table*}


\subsection{H254 in the Hertzsprung--Russell diagram}

\begin{figure}
\centering
\includegraphics[width=9cm]{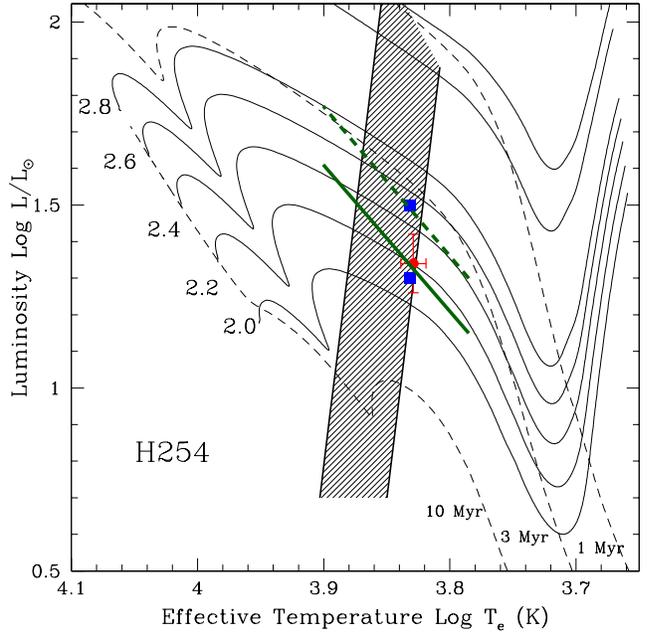}
\caption{The Hertzsprung--Russell diagram for H254: solid red circle
  represents the position of H254 obtained by using the
  spectroscopic temperature and the luminosity derived from the BW
  analysis. Solid blue squares show the best fit models of
  Fig.~\ref{Jagoda}. The dark vertical band represents
  the instability strip from \citet{marconi98}, while evolutionary
  tracks for different values of the mass 
  (solid lines) and the isochrones for different values of the ages
  (dashed lines) are obtained from \citet{tog11}. Finally, solid and
  dashed green lines represent the locus of constant period (equal
  to 7.406$^{-1}$d) for fundamental and first overtone mode,
  respectively.}   
\label{fig-hr}
\end{figure}

Using the distance obtained in the
previous section, the apparent magnitude of H254 and our estimate for 
the absorption, as well as the assumed BC, we can evaluate the
intrinsic luminosity of H254. Taking into account the errors
associated to the various quantities, we obtain:
$logL/L_{\odot}=1.34\pm 0.08$ dex.
The location of the star in  the  Hertzsprung--Russell diagram 
(filled red circle in Fig.~\ref{fig-hr} with the associated error bars), is evaluated combining this luminosity estimate
with our spectroscopic measurement of the effective temperature, namely
$T_{\rm eff}$=6750$\pm$150 K. 
We notice that the evaluated position of H254  is close to the red edge of the instability strip. In the same
  plot, the filled blue squares mark the location of the two best fit
  models, used to define the pulsational mode of H254 \citep{jagoda03,jagoda05}. Empirical data
  seems to suggest that the lower luminosity model with
  $logL/L_{\odot}=1.3$ should be preferred.

Using the recent results by \citet{tog11}, we overplot in Fig.~\ref{fig-hr} the
evolutionary tracks for selected masses, ranging from 2.0 $M_\odot$ to 2.8
$M_\odot$, and three isochrones at 1, 3 and 10 Myr. From this analysis
we can conclude that H254 has a mass of about 2.2 $M_\odot$ and an
age of 5$\pm$1 Myr. Comparing these results with
those obtained in Paper I, we can conclude that H254 has almost
the same mass estimated in our previous work, but it results
older in the present analysis. 
Additionally, we note that the radius estimated from the \citet{tog11} 
2.2 $M_\odot$ evolutionary track at the temperature of H254 is 3.4 R$_\odot$, in very
good agreement with the CORS result.  


The solid and dashed magenta lines in Fig.~\ref{fig-hr} represent the
loci of constant fundamental and first overtone period
respectively, predicted by linear nonadiabatic models \citep[see][for details]{marconi98},
with  period equal to the observed one.
We notice that, even taking into account the errors on the star
luminosity and effective temperature, the fundamental mode is
favoured.
 

\section{Conclusions}

In this paper we present new photometric and radial velocity data
for the PMS $\delta$ Sct star H254, member of the young cluster 
IC 348. The photometric  light curves were secured in the Johnson-Cousins
$V,R_C,I_C$ bands using the Loiano and Asiago  telescopes. The radial
velocity data was acquired by means of the SARG@TNG spectrograph.  
Some of the high-resolution spectra were specifically acquired with SARG  to estimate 
stellar parameters and the chemical composition of the star, obtaining: T$_{\rm
  eff}$\,=\,6750\,$\pm$\,150~K; $\log g$\,=\,4.1\,$\pm$\,0.4 dex; 
[Fe/H]=-0.07$\pm$0.12 dex.  We note that the T$_{\rm  eff}$ derived
here is cooler by more than 400 K with respect to previous literature
results based on low-resolution spectroscopy.  
Photometric and spectroscopic data were  
used to estimate the total absorption in the $V$ band
A$_{\rm V}$=2.06$\pm$0.05 mag, a value that is in agreement with previous
estimates. 

To the aim of estimating the radius of H254 by applying a
Baade-Wesselink technique, we have first ascertained that H254 
pulsate in a radial mode by adopting the technique of the difference
in phase and amplitude between different photometric bands and radial
velocities. As a result H254 was confirmed to pulsate in a radial mode. 

Using the photometric and spectroscopic data, we have applied the CORS
realization of the Baade--Wesselink method in the form developed by
\citet{rip97}, obtaining a value for the linear radius of H254 equal
to 3.3$\pm$0.7 $R_\odot$. This quantity was used to measure the
distance of the target star and, in turn, of the host cluster IC
348, obtaining a final value of  273$\pm$23 pc. This estimate is in agreement within the
errors with the results derived from the  Period--Luminosity relation for
$\delta$--Scuti stars by \citet{mcn11} and those coming from the
Hipparcos parallaxes by \citet{sch99} as well as with our own recalculation based
on the revised Hipparcos parallaxes by \citet{vanleeuwen07}. We are
only marginally in agreement with the larger value of the distance,
i.e. 310-320 pc found by \citet{her98} and \citet{dezeeuw99}. 

Finally, we derived the luminosity of H254 and studied its position
in the Hertzsprung--Russell diagram. From this analysis it results
that this $\delta$--Scuti occupies a position close to the red edge of
the instability strip, pulsates in the fundamental mode, 
has a mass of about 2.2 $M_\odot$ and an age
of 5$\pm$1 Myr, older than previous estimates.

\section*{Acknowledgments}

We thank our anonymous Referee for his/her very helpful comments 
that helped in improving the paper.
It is a pleasure to thank M.I. Moretti for a critical  reading of the
manuscript. 
This research has made use of the SIMBAD database, operated at CDS, Strasbourg, 
France. 

\end{document}